\documentclass[useAMS,usenatbib,paper]{mn2e}
\usepackage{times,epsfig,psfig,natbib,amssymb,amsmath,graphicx}
\usepackage[bookmarks,bookmarksnumbered,colorlinks=true, citecolor=blue, linkcolor=black]{hyperref}
\usepackage{color,natbib}
\usepackage[normalem]{ulem}

\title[Spectropolarimetry of a Radio-loud NLSy1]{Radio-loud
  Narrow Line Seyfert 1 under a different perspective:\\ a revised
  black hole mass estimate from optical spectropolarimetry}
\author[Baldi et al.]{Ranieri
  D. Baldi$^{1,2}$\thanks{E-mail:R.Baldi@soton.ac.uk}, Alessandro
  Capetti$^{3}$, Andrew Robinson$^{4}$, Ari Laor$^{2}$, and Ehud
  Behar$^{2}$\\ $^{1}$Department of Physics and Astronomy, The
  University, Southampton SO17 1BJ, UK\\ $^{2}$Physics Department, The
  Technion, 32000, Haifa, Israel\\$^{3}$INAF-Osservatorio Astrofisico
  di Torino, Strada Osservatorio 20, I-10025, Pino Torinese,
  Italy\\$^{4}$School of Physics and Astronomy, Rochester Institute of
  Technology, 84 Lomb Memorial Drive, Rochester, NY 14623-5603, USA}

\begin{document}



\maketitle

\label{firstpage}

\begin{abstract}

  Several studies indicate that radio-loud (RL) Active Galactic Nuclei (AGN)
  are produced only by the most massive black holes (BH), $M_{\rm BH} \sim
  10^8$-$10^{10} M_\odot$. This idea has been challenged by the discovery of
  RL Narrow Line Seyfert 1 (RL~NLSy1), having estimated masses of $M_{\rm
    BH}$$\sim$$10^6$-$10^7$ M$_\odot$. However, these low $M_{\rm BH}$
  estimates might be due to projection effects. Spectropolarimetry allows us
  to test this possibility by looking at RL~NLSy1s under a different
  perspective, i.e., from the viewing angle of the scattering material. We
  here report the results of a pilot study of VLT spectropolarimetric
  observations of the RL~NLSy1 PKS~2004-447. Its polarization properties are
  remarkably well reproduced by models in which the scattering occurs in an
  equatorial structure surrounding its broad line region, seen close to
  face-on. In particular, we detect a polarized H$\alpha$ line with a width of
  $\sim$ 9,000 km s$^{-1}$, $\sim 6$ times broader than the width seen in
  direct light. This corresponds to a revised estimate of $M_{\rm
    BH}$$\sim$$6\times10^8$ M$_\odot$, well within the typical range of RL
  AGN.

\end{abstract}


\section{Introduction}

Narrow line Seyfert 1 (NLSy1) are AGN in which the width of the broad emission
lines is smaller than 2,000 km s$^{-1}$ and showing a [O~III]/H$\beta$ ratio
smaller than 3 \citep{osterbrock85}. The estimated masses of their black holes
(BH), based on the FWHM of the BLR lines and the continuum luminosity
(e.g. \citealt{Kaspi00}), are typically $M_{\rm BH}$$\sim$$10^6$-$10^7$
M$_\odot$. A recent surprising result is the discovery of a sub-population of
NLSy1s with characteristics typical of radio-loud (RL) AGN (e.g.,
\citealt{zhou06,yuan08,dammando12,foschini12,angelakis15}). In contrast,
classical RL AGN are invariably associated with the most massive BH, with
$M_{\rm BH} > 10^{8} M_\odot$ (e.g. \citealt{laor00,baldi10b,chiaberge11}). This
result suggests two possible interpretations: RL~NLSy1s correspond to a
fundamentally different mechanism for the formation of relativistic jets
(e.g. \citealt{komossa06,gu15}). Alternatively, the $M_{\rm BH}$ estimates in
RL~NLSy1s are underestimated.

One possibility to account for their ``narrow'' broad Balmer lines is that the
BLR clouds in RL~NLSy1s are mostly confined in a rotating disk. When such a
disk is seen at a small angle with respect to our line of sight, projection
effects reduce the observed line width leading to an under estimate of the BH
mass values. Several indications favor a pole-on orientation for RL~NLSy1,
including rapid variability, one-sided radio jets, and extreme $\gamma$-rays
luminosity, all requiring high Doppler boosting typical of relativistic jets
observed at a small angle from their axis (e.g.,
\citealt{komossa06,doi07,komossa15}). The presence of a dependence of the BLR line
width with orientation is already well established for RL AGN (e.g.,
\citealt{wills86,Fine11,runnoe13}), in the sense that objects observed at
smaller angles from their jet axis show narrower lines. RL~NLSy1s might be
extreme examples of such an effect.

Spectropolarimetry is a unique tool to probe the geometry of
AGN. Among the many results obtained with this technique there is the
Unified Model, proposed to account for the polarized broad lines seen
in objects showing only narrow lines in direct light
\citep{miller83,antonucci83,antonucci85a}. This is due to the
scattering of the light of a hidden BLR into our line of sight,
leading to a polarized emission. Also in the case of NLSy1s
spectropolarimetry enables us to observe their BLR under a different
perspective. For example, a scattering region located in the AGN
equatorial plane will see the full rotation amplitude of the BLR and
the light reflected (and thence polarized) will show the intrinsic
value of the BLR line width.  The presence of scattering
material in the equatorial plane is supported by the polarization
properties of the majority of type 1 AGN, where the polarization is
parallel to the radio axis (while in type 2 is usually perpendicular)
and the polarization angle rotates within the broad line. This
behavior is well reproduced by a geometry in which the scatterers are
mainly located in an annulus immediately surrounding the BLR
\citep{smith04,smith05}.

Several spectropolarimetric studies of NLSy1s has been performed
(e.g. \citealt{goodrich89,breeveld98,kay99,robinson11}) but none of
them included RL objects of this class.

We here present the results of a pilot study of VLT spectropolarimetric
observations of a RL~NLSy1, namely PKS~2004-447, at $z=0.240$. We selected it
because of its detection (and variability) in $\gamma$-rays by {\it Fermi}
\citep{abdo09,calderone11}, the clearest evidence for the presence of a RL
nucleus. Among the few RL~NLSy1s with a $\gamma$-rays detection, PKS~2004-447
is the only source in the Southern sky at sufficiently low redshift 
that the H$\alpha$ line falls within an optical spectrum.

PKS~2004-447 is a compact steep-spectrum radio source, with a core-jet
morphology on pc scales, \citep{schulz15} and its broadband spectral energy
distribution (SED) shows the double-humped shape typical of blazars
\citep{gallo06,orienti15}. The FWHM of its broad H$\beta$ line is 1447
km~s$^{-1}$ \citep{oshlack01}, within the upper limit considered to define
NLSy1s; the corresponding BH mass is $\sim 5 \times 10^6 M_\odot$.  The
strength of the Fe II lines is low (with an equivalent width for the whole
Fe~II complex in the region 5050-5450 \AA\ smaller than 10 \AA) with respect
to what is usually found in NLSy1s \citep{osterbrock87,boroson92}; conversely,
the ratio [O~III]/H$\beta =1.6$ conforms with typical values of NLSy1.
\citeauthor{oshlack01} also estimated a radio-loudness $R = f_{\rm 4.85
  \,\,GHz}/f_B \sim 1710 - 6320$ (depending on the value used for the optical
magnitude) well above the threshold for RL AGN ($R >$10,
\citealt{kellerman89}).
\begin{figure*}
\label{fig}
\centering  
\includegraphics[width=13.cm,angle=90]{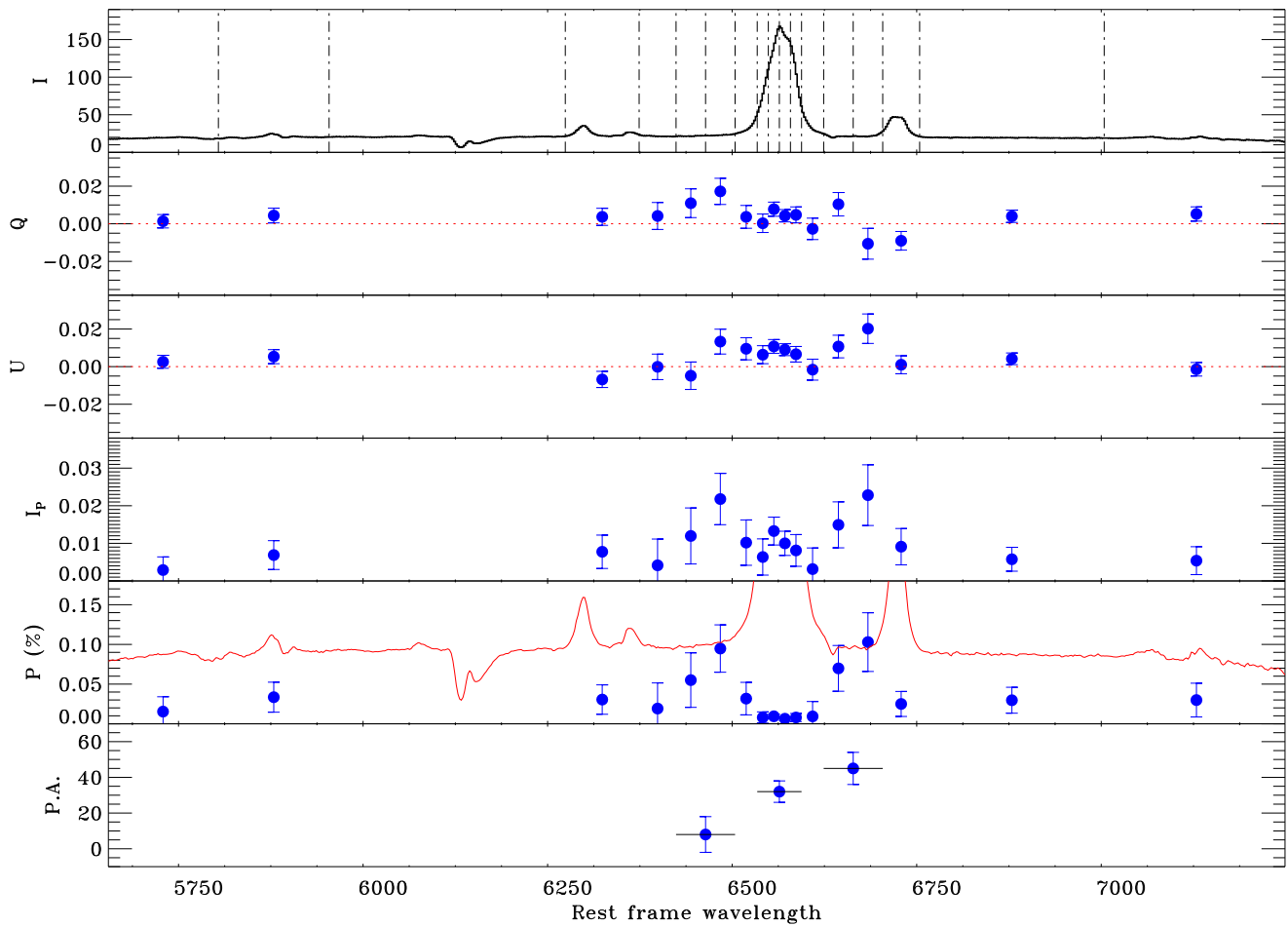}
\caption{Top three panels: Stokes parameters of the PKS~2004-447
  spectrum. Wavelengths are shown in rest frame in \AA, fluxes are in
  arbitrary units because no flux calibration was performed. The vertical
  dashed lines mark the boundaries of the regions used for the spectral
  rebinning. Bottom three panels: polarized flux, percentage of polarization
  (with overplotted, in red, the total intensity spectrum), and polarization
  position angle. The PA is reported only in the regions of reliable
  polarization, see text for details. The region around $\sim$ 6100 \AA\ has
  been flagged due to the residuals related to the presence of a telluric
  absorption feature and of bright sky lines.}
\end{figure*}

A cosmology with $H_0 = 72$ km s$^{-1}$ Mpc$^{-1}$, $\Omega_m = 0.30$, and
$\Omega_\Lambda = 0.70$ is assumed, corresponding to a luminosity distance for
PKS~2004-447 of 1.17 Gpc.

\section{Observations and data analysis}
\label{observations}

PKS~2004-447 was observed in spectropolarimetric mode with the FORS2 mounted
at the UT1 telescope of the 8.2-m ESO Very Large Telescope (VLT) on 2015 May
11 and 23 with a seeing better than $<$1\arcsec. The observations were
obtained with the GRISM-300I, providing a spectral resolution of $R\sim$660
at 8600 \AA, used in combination with the blocking filter OG590. The
1$\arcsec$ wide slit was oriented along the parallactic angle and the
multi-object spectropolarimetry (PMOS) observation was performed with a 2048
$\times$ 2048 pixel CCD with a spatial resolution of 0.126 \arcsec/pixel.

The measurements at two retarder angles are sufficient to calculate the linear
Stokes parameters. However, in order to reduce instrumental issues
\citep{patat06} we used four half-wave plate angles (0, 22.5, 45 and 67.5
degrees) for our observations. This strategy was repeated four times in
separated observing blocks. Each block consists of four spectra (split in two
exposures for cosmic rays removal) with an exposure time of 10.25 minutes,
with the half-plate oriented at the four position angles. Since the
observation of the first night failed, we used only three observing blocks,
for a total on-source time of $\sim$2 hours.

The data were analyzed by using the standard VLT pipeline with the ESO Reflex
workflow \citep{hook08}. The frames were first bias-subtracted and then
flat-fielded using lamp flats.  After removing cosmic ray events, the
different exposures for the four half-wave plate positions were combined for
each observing block.  One-dimensional spectra, with identical aperture
widths for the ordinary (o) and extraordinary (e) rays for each of the four
half-wave plate orientations, were then extracted (from 12 pixels
corresponding to 1.5\arcsec). The wavelength calibration was performed using
standard arc lamps.  The o- and e-rays were resampled to the same linear
spectral dispersion and combined (by including data for all three observing
blocks) to determine the three Stokes parameters I, Q, and U (as well as the
associated uncertainties) using the procedures described by \citet{cohen95}
and \citet{vernet01}. No flux calibration was performed.

\section{Results}

The total intensity spectrum is shown in the top panel of Figure~\ref{fig}.
It shows a broad H$\alpha$ line, blended with [N~II]$\lambda\lambda$6548,84
doublet, with a FWHM$\sim$1500 km s$^{-1}$, consistent with the H$\beta$ width
measured by \citet{oshlack01}. Fainter emission lines are also seen, namely
the [O~I]$\lambda\lambda$6300,64 and [S~II]$\lambda\lambda$λ6716,31 doublets,
superposed onto a featureless continuum. The signal-to-noise ratio of the
total intensity spectrum is $\sim$150 per pixel on the continuum and it
reaches $\sim$500 on the H$\alpha$.

Since the polarization of the target turns out to be extremely low (see below
for the details) the spectra were rebinned in order to obtain, where possible,
statistically significant polarization measurements. The bin sizes used are
broader in the continuum far from the spectral regions of interest (i.e., the
broad H$\alpha$ line), where bin sizes of either 150 and 250 \AA, were used. A
finer binning (15 \AA) is maintained across the H$\alpha$, while the
  integration regions are between 30 and 50\AA\ on its wings. In Fig.~\ref{fig}
we present the resulting Stokes parameters $Q$ and $U$, the polarized spectrum
$I_P =(Q^{2} + U^{2})^{1/2}$, the corresponding degree of polarization $P =
I_P/I$, and the polarization position angle, PA (measured counterclockwise
starting from North). The PA is plotted only in the regions of significant
polarization.

The overall polarization level of this source (integrated between 5404
and 7254 \AA) is consistent with a null value, namely $P= 0.03 \pm
0.02$\%, despite the very high accuracy that can be reached with our
data. The same result is obtained for the continuum emission. We
measured it in two spectral regions $\sim$ $\pm$ 500 \AA\ far from the
broad H$\alpha$ (covering 5404-5904 and 6754-7254 \AA, respectively)
failing to detect any significant polarization.

The BLR seen in direct light does show a polarization of 0.0076 $\pm$ 0.0024
\% integrated over the wavelength range 6534-6594 \AA, with a polarization
position angle of PA = $32^\circ \pm 6^\circ$ . While its polarization has a
statistical significance of $\sim$ 3$\sigma$, we cannot exclude that we are
seeing a minimal residual instrumental polarization. Certainly, this
measurement can be adopted as a firm upper limit to any instrumental effect.

The only reliable polarized signal found in PKS~2004-447 is seen on
both sides of the broad H$\alpha$.  We measure a degree of
polarization of $P$ = 0.066 $\pm$ 0.023\% at PA= $8^\circ \pm
10^\circ$ on the blue side (for 6424 $< \lambda < $ 6504 \AA) and $P$
= 0.070 $\pm$ 0.023\% at PA = $45^\circ \pm 9^\circ$ on the red side
(for 6624 $< \lambda < $ 6704 \AA).

\section{Discussion}
\label{discussion}

The spectropolarimetric observations show a broader polarized H$\alpha$
with respect to what is seen in direct light. This is what is expected if we
are seeing the H$\alpha$ emission from a disk-like BLR reflected from a
co-planar circumnuclear scattering region. Due to the location of the
scatterers, they see the full BLR velocity field; thus the amplitude of the
scattered (and thence polarized) broad line reflects the genuine gas motions
close to the central BH. This confirms our suggestion that projection plays a
fundamental role in RL~NLSy1.

The polarization behavior of PKS~2004-447 is in remarkable agreement with the
results of the models of presented by \citet{smith05}. They simulated the
polarization properties of a rotating line-emitting disc (the BLR) surrounded
by a coplanar scattering region. The main predictions are i) the presence of
polarization peaks in the line wings, ii) a rotation of the polarization PA
across the BLR, iii) a higher polarization in the line with respect to the
continuum, all effects seen in the observed source.

In the equatorial scattering models, the polarization level decreases
for sources seen at lower inclination. This is due to the cancellation
between the polarization vectors produced in the different portions of
the scattering region, which is more efficient for low inclination
angles (a null $P$ value is predicted for an exact face-on
orientation). Indeed, the measured percentage of polarization
($\sim$0.07\%) is an order of magnitude smaller than usually observed
in type I Seyfert \citep{smith02}, as expected for a source seen close
to face-on. As discussed above, this is likely to be the case for our
target. 

Another signature of equatorial scattering is the angle swing across
the H$\alpha$ line. The PA of the continuum polarization defines the
reference value, parallel to the projection of the symmetry
axis. Across the polarized broad line, the PA rotates in opposite
directions in the blue and red wings, with the largest deviations
occurring just inside the peaks of polarization percentage. This
feature is seen in PKS~2004-447 with a rotation of $37^\circ \pm
13^\circ$ from the blue to the red H$\alpha$ wings. However, the non
detection of continuum polarization (that is indeed expected to be
lower than on the line wings) does not allow us to trace fully the PA
changes. Furthermore, the actual rotation across the line might be
even larger considering the heavy smoothing of our data. Nonetheless,
the observed PA swing is already larger than usually measured in
Seyferts~Is and indeed the models indicate that such effect is
amplified for small inclinations, as in the case for our target.

With respect to the model predictions we find a possible discrepancy, related
to the relative PA of the BLR polarization and the radio jet that are expected
to be parallel to each other. The BLR polarization angle, averaged over the
two wings, is $27^\circ \pm 7^\circ$ while the radio jet is elongated in a
general EW direction \citep{orienti15}. However, the jet is strongly bent,
starting along PA$\sim -90^\circ$ out to $\sim$15 pc, where it curves
Northward by $\sim 45^\circ$.  Large bends are commonly observed in blazar
radio jets, often also associated with temporal changes of their orientation
\citep{lister13}; this is due to projection effects that, in objects seen
pole-on, amplify the intrinsic variation of the jet position angles. For such
objects it is difficult to establish in which direction the jet is actually
pointing. Furthermore, \citet{smith04} noticed that Seyfert type 1 galaxies
often show a similar perpendicular inclination between the radio and BLR
position angles. They proposed that an asymmetry of the density distribution
of the scattering region might account for such a configuration.  For these
reasons, the apparent offset between the position angles of the BLR
polarization and the jet orientation found in PKS~2004-447 should not be
considered as a strong shortcoming of the scattering model proposed.

According to the standard equatorial model of \citet{smith05} (see
their Figure 4) the separation between the peaks in the percentage of
polarization corresponds to the intrinsic (i.e. corrected for
projection) width of the broad H$\alpha$.  In our case, although, due
to the low polarization level, this measurement has a relatively large
uncertainty.  In these spectral regions the bin sizes are of 30-40
\AA; by adopting these values as errors for the location of the two
polarization peaks, they turn out to be separated by 200 $\pm$ 50 \AA,
corresponding to a velocity width of 9100 $\pm$ 2300 km s$^{-1}$, a
value within the range covered by the broad Balmer lines in RL AGN
\citep{wills86,runnoe13}.

The broadening with respect to what is seen in direct light is
dramatic. As a consequence, the BH mass estimate is increased by a
factor $\sim$30 with respect to the estimate obtained from the total
intensity spectrum. Given the continuum luminosity of $L_{5100} \sim
1.25 \times 10^{44}$ erg s$^{-1}$ cm$^{-2}$ \AA$^{-1}$
\citep{gallo06}, the corresponding BLR size is $\sim 38$ light days
\citep{bentz13}. By adopting the width of the polarized broad
  H$\alpha$,\footnote{We assumed a geometric factor $f=1$ in the
  virial formula to account for the edge-on location of the
  scatterers.} we derive a mass of $M_{\rm BH} \sim 6\times10^8
M_\odot$. This revised estimate falls in the range of the BH masses
typical of RL AGN \citep{chiaberge11}. 

\citet{calderone13} reached a similar conclusion from a method to
estimate black hole masses which relies on the modeling of optical and
UV data of of 23 RL-NLSy1 with an accretion disk spectrum. Their study
returns values of $M_{\rm BH}$ consistently larger than
$\gtrsim$10$^{8}$ M$_{\odot}$.

The results obtained from our study lead us to consider the further issue of
why RL~NLSy1s are so rare. In principle, flat spectrum radio-quasars (FSRQ),
objects known to be observed at small inclinations, should also show narrow
BLR profiles. While indeed the FWHM of the broad emission lines in FSRQ
is reduced with respect to the general population of QSOs, only a small
minority of them have a width $<$ 2000 km s$^{-1}$ \citep{shaw12}. This
implies that projection effects are not as effective in these sources; this is
expected when non-rotational motions, such as e.g. those associated with an
outflow contribute significantly to the BLR velocity field or when the
  BLR is characterized by a large vertical thickness. It appears that a
combination of face-on inclination and a BLR dominated by rotation (in the
Balmer lines) are probably required to produce a RL~NLSy1.

The dominance of rotation was proposed by \citet{eracleous03} to account for
the BLR properties of another class of objects, the so-called
``double-humped'' AGN (DH-AGN) in which the broad line profiles show two well
separated emission peaks; DH-AGN constitute $\sim$20\% of RL-AGN for
$z<0.4$. In most cases the ``double-humped'' profiles can be described very
well by models attributing the emission to a flattened rotating BLR (for a
review see \citealt{eracleous09}). By extrapolating the results obtained for
PKS~2004-447, we speculate that RL~NLSy1s are the result of the rare
combination of an intrinsically DH-AGN that is seen nearly face-on.

Regardless of the proposed identification of the parent population (i.e., of
the objects with the same intrinsic properties but observed at larger
inclinations), the expected broadening of the BLR for RL~NLSy1s seen at large
angles is an effect that should be included in the search for their
mis-oriented counterparts (see, e.g., \citealt{berton15}).

\section{Summary and conclusions}

The spectropolarimetric observations of the RL~NLSy1s PKS~2004-447 show the
presence of two peaks of polarized flux separated by $\sim$ 250\AA\, flanking
the H$\alpha$  seen in the total intensity spectrum. The overall polarization
properties of this source are remarkably well reproduced as due to the
scattering of the BLR emission from an equatorial structure seen at very small
inclination from its axis. In particular this model predicts the extremely low
level of polarization ($\sim$0.07\%) and the change of position angle across
the broad H$\alpha$. In this situation the scattering material, from its vantage point,
sees the full velocity field of the BLR, removing the effects of
projection. The separation between the polarization peaks is a direct measure
of its intrinsic width, that results in $\sim$9,000 km s$^{-1}$. By adopting
the standard scaling relations we derive a black hole mass of $M_{\rm BH} \sim
6\times10^8 M_\odot$. This revised estimate strengthens the conclusion that RL
AGN are only associated with very massive black holes ($\gtrsim$10$^{8}$
M$_{\odot}$).

It must be stressed that our results do not apply to the class of NLSy1s as a
whole. In fact, we are interpreting the observations of an object that was
selected among the small sub-class of RL sources, while the vast majority of
NLSy1s are radio-quiet (RQ). In particular, the large BH mass estimated for
PKS~2004-447 does not necessarily conflict with the suggestion, based on
various properties of NLSy1, that they are generally associated with small BHs
accreting at high rates (e.g., \citealt{valencia12}). Indeed, previous
spectropolarimetric studies of RQ~NLSy1s (e.g., \citealt{goodrich89}) do not
show a H$\alpha$ broadening in polarized light. These latter results
  apparently contrast with the suggestion by \citet{decarli08} that the
  properties of NLSy1 can be accounted for with a preferential face-on
  orientation. Furthermore, although PKS~2004-447 qualifies as a NLSy1s
based on its main optical properties (broad emission line width and
[O~III]/H$\beta$ ratio), it shows rather different characteristics (such as
its broad-band SED and the strength of the Fe~II lines) with respect to what
is typical for NLSy1. We concur with \citet{oshlack01} who argue that since
NLSy1s are defined by phenomenological measurements, rather than a specific
physical model, it is possible that more than one physical mechanism will
produce the defining parameters. In particular, the low widths of their broad
lines is not generally due to projection. Our spectropolarimetry results
strongly suggest that this is, in fact, the case for at least one RL NLSy1,
although, clearly, more spectropolarimetric observations will be required to
determine if this conclusion holds for the RL sub-population in general.

On the other hand, not all RL AGN seen at small inclinations have small 
  broad emission line widths; indeed only a minority of them show values
smaller than the threshold of 2,000 km s$^{-1}$ required for a classification
as NLSy1. This indicates that a further ingredient, other than orientation, is
needed to produce a RL~NLSy1. In particular, the dependence of the  broad
  emission line width with orientation is maximized when its velocity field
is dominated by rotation. We speculate that the parent population of RL~NLSy1s
is represented by sub-class of the so called ``double humped'' AGN; in these
objects the lines profiles can be reproduced accurately by models attributing
the emission to a flattened rotating BLR. We argued that RL~NLSy1s are the
result of the rare combination of an intrinsically DH-AGN that is seen nearly
face-on.

Our results also suggest that orientation effects might play an
important role in the BH estimates based on the virial method not only
for NLSy1 but more in general for AGN and particularly for those seen
close to face-on. For example, in FSRQ the median angle formed by the
jet with the line of sight is $\sim3^\circ$ \citep{savolainen10} and
this reduces the rotation amplitude by a factor of $\sim$ 20 with
respect to its intrinsic value in case of a disk-like
geometry. Furthermore, since disk winds seem to be important for gas
dynamics and ubiquitous in AGN (e.g. \citealt{veilleux05}), any
outflowing component might represent the dominant source of observed
line broadening in these AGN. These factors cast doubts on the
reliability of virial estimates of their BH masses.

\section*{Acknowledgments}

RDB is grateful to S. Raimundo for the discussion which inspired this
work.  We thank the referee for constructive comments/suggestions to
improve the manuscript.  EB received funding from the European Unions
Horizon 2020 research and innovation programme under the Marie
Sklodowska-Curie grant agreement no. 655324, and from the I-CORE
program of the Planning and Budgeting Committee (grant number
1937/12).

\bibliography{./my.bib}

\end{document}